\documentclass[shortnote,twocolumn]{jpsj2}

\def\gsim{\mathop {\vtop {\ialign {##\crcr 
$\hfil \displaystyle {>}\hfil $\crcr \noalign {\kern1pt \nointerlineskip
 } 
$\,\sim$ \crcr \noalign {\kern1pt}}}}\limits}
\def\lsim{\mathop {\vtop {\ialign {##\crcr 
$\hfil \displaystyle {<}\hfil $\crcr \noalign {\kern1pt \nointerlineskip
 } 
$\,\,\sim$ \crcr \noalign {\kern1pt}}}}\limits}

\title{Antiferromagnetic Exchange Interaction between Electrons on Degenerate LUMOs in Benzene Dianion }

\author{Hiroyasu {\sc Matsuura}$^{1,2}$, Kazumasa {\sc Miyake}$^{2}$, and Hidetoshi {\sc Fukuyama}$^{3}$} 

\inst{$^{1}$Department of Physics, University of Tokyo, 7-3-1 Hongo, Bunkyo, Tokyo 113-0033, Japan\\
$^{2}$Division of Materials Physics, Department of Materials Engineering Science \\
Graduate School of Engineering Science, Osaka University, Toyonaka, 
Osaka 560-8531, Japan\\
$^{3}$Faculty of Science and Research Institute for Science and Technology (RIST)
Tokyo University of Science\\ 
1-3, Kagurazaka, Shinjuku-ku, Tokyo 162-8601, Japan
}

\abst{This document briefly provides instructions on how to prepare your manuscript in \LaTeX\ format. As regards general instructions for preparing manuscripts, please refer to ``Instructions for Preparation of Manuscript", which is available at our Web site (http://jpsj.ipap.jp/).}

\kword{Benzene dianion (Bz$^{2-}$), Antiferromagnetic exchange coupling}

\begin{document}
\maketitle

Many aromatic hydrocarbons have doubly or multiply degenerate lowest unoccupied molecular orbitals (LUMOs) consisting of 2$p\pi$-orbitals.
The exchange interaction between electrons on LUMOs in these dinegative ions tends to be considered as ferromagnetic (the spin-triplet state) by the Hund's rule coupling.  
Indeed, the triplet ground state has been observed for the dinegative ions of triphenylbenzene, decacyclene, and so on~\cite{Murrell}.
Benzene (Bz) has doubly degenerate LUMOs.
Theoretically, the electronic state of Bz dianion (Bz$^{2-}$) has been studied by quantum chemical calculations~\cite{Jesse,Nakayama}.
It has been reported that the ground state is the triplet state, and the energy difference between the singlet state and triplet state is about $0.1$ eV~\cite{Jesse} or about $1$ eV~\cite{Nakayama}.  
However, since the many body effects have not been fully treated and the energy difference is tiny, it has not yet been determined whether the ground state is the spin triplet state or the spin singlet state.
In addition, the ground state of Bz$^{2-}$ has not been experimentally studied. 

In the present paper, we discuss the ground state of Bz$^{2-}$ on the basis of the numerical diagonalization method of an effective model of $\pi$ orbitals.
Then, we estimate the amplitude of the exchange coupling between two electrons on LUMOs.
As a result, we find that the ground state becomes the spin singlet state, and  the exchange coupling between LUMOs can be antiferromagnetic.

The model Hamiltonian for Bz$^{2-}$ is given as follows:
\begin{eqnarray}
\mathcal{H} =\mathcal{H}_0 + \mathcal{H}_{1}, 
\end{eqnarray}
where
\begin{eqnarray}
\mathcal{H}_0 &=& -t_{\pi} \sum_{i=1}^{6}\sum_{\sigma}\left[ \chi_{i\sigma}^{\dag} \chi_{i+1\sigma} + {\rm h.c.} \right], \\ 
\mathcal{H}_{1} &=& U \sum_{i=1}^{6}n_{i\uparrow }n_{i\downarrow } + (U^\prime - J)\sum_{i=1}^{6}\sum _\sigma n_{i\sigma}n_{i+1\sigma} \nonumber \\
  && + U^\prime \sum_{i=1}^{6}\sum _{\sigma \neq \sigma^\prime} n_{i\sigma}n_{i+1\sigma^\prime}\nonumber \\ 
   && + J\sum_{i=1}^{6}\big[ \chi_{i\uparrow}^{\dag}\chi_{i+1\uparrow}\chi_{i+1\downarrow }^{\dag}\chi_{i\downarrow }    \nonumber \\ 
   && +\chi_{i\uparrow}^{\dag}\chi_{i+1\uparrow}\chi_{i\downarrow }^{\dag} \chi_{i+1\downarrow}+{\rm h.c.} \big] ,
\end{eqnarray}
where $\chi$'s are atomic 2$p\pi$ orbitals of Bz; $t_{\pi}$, $U$, $U'$, and $J$ represent the transfer integral, the intra- and inter-orbital Coulomb interaction and the ferromagnetic exchange coupling, respectively.
This effective model is an extended Hubbard model ($t$-$U$-$V$-$J$ model).
It is well known that the exchange coupling, $J$, plays a fundamental role in the property of the ground state~\cite{Hirsch,Tang}.

The amplitude of $U$ and $t_{\pi}$ have been estimated by the quantum chemical calculation: $U \sim 17$ eV and $t_{\pi} \sim 2.5$eV~\cite{McWeeny,Benzene}.
Thus, we choose $U/t_{\pi}=7.0$.
The other parameters have also been estimated by the same quantum chemical calculations: $U^{\prime} \sim 9$ eV ($U^{\prime}/t_{\pi} \sim 3.6$) and $J \sim 1$ eV ($J/t_{\pi} \sim 0.4$).
However, since we neglect many other interactions (for example, the second nearest Coulomb interactions), we regard $U^{\prime}$ and $J$ as variable parameters. 

By diagonalizing the kinetic energy term $\mathcal{H}_{0}$, six molecular orbitals, $\phi_{1} \sim \phi_{6}$, are given as follows~\cite{Benzene}:
\begin{eqnarray}
\phi_{1}&=&{1\over \sqrt{6}}(\chi_{1}+\chi_{2}+\chi_{3}+
\chi_{4}+\chi_{5}+\chi_{6}), \label{eq:1a}
\\
\phi_{2}&=&{1\over 2\sqrt{3}}(2\chi_{1}+\chi_{2}-\chi_{3}
-2\chi_{4}-\chi_{5}+\chi_{6}),
\label{eq:1b}
\\
\phi_{3}&=&{1\over 2}(\chi_{2}+\chi_{3}-
\chi_{5}-\chi_{6}),
\label{eq:1c}
\\
\phi_{4}&=&{1\over 2}(\chi_{2}-\chi_{3}+
\chi_{5}-\chi_{6}),
\label{eq:1d}
\\
\phi_{5}&=&{1\over 2\sqrt{3}}(2\chi_{1}-\chi_{2}-\chi_{3}
+2\chi_{4}-\chi_{5}-\chi_{6}),
\label{eq:1e}
\\
\phi_{6}&=&{1\over \sqrt{6}}(\chi_{1}-\chi_{2}+\chi_{3}-
\chi_{4}+\chi_{5}-\chi_{6}).
\label{eq:1f}
\end{eqnarray}
The eigen energy increases from $\phi_{1}$ to $\phi_{6}$, while $\phi_{2}$ ($\phi_{4}$) and $\phi_{3}$ ($\phi_{5}$) are degenerate.
We call $\phi_{4}$ ($\phi_{5}$) LUMO.

\begin{figure}[h]
\begin{center}
\rotatebox{0}{\includegraphics[width=1\linewidth]{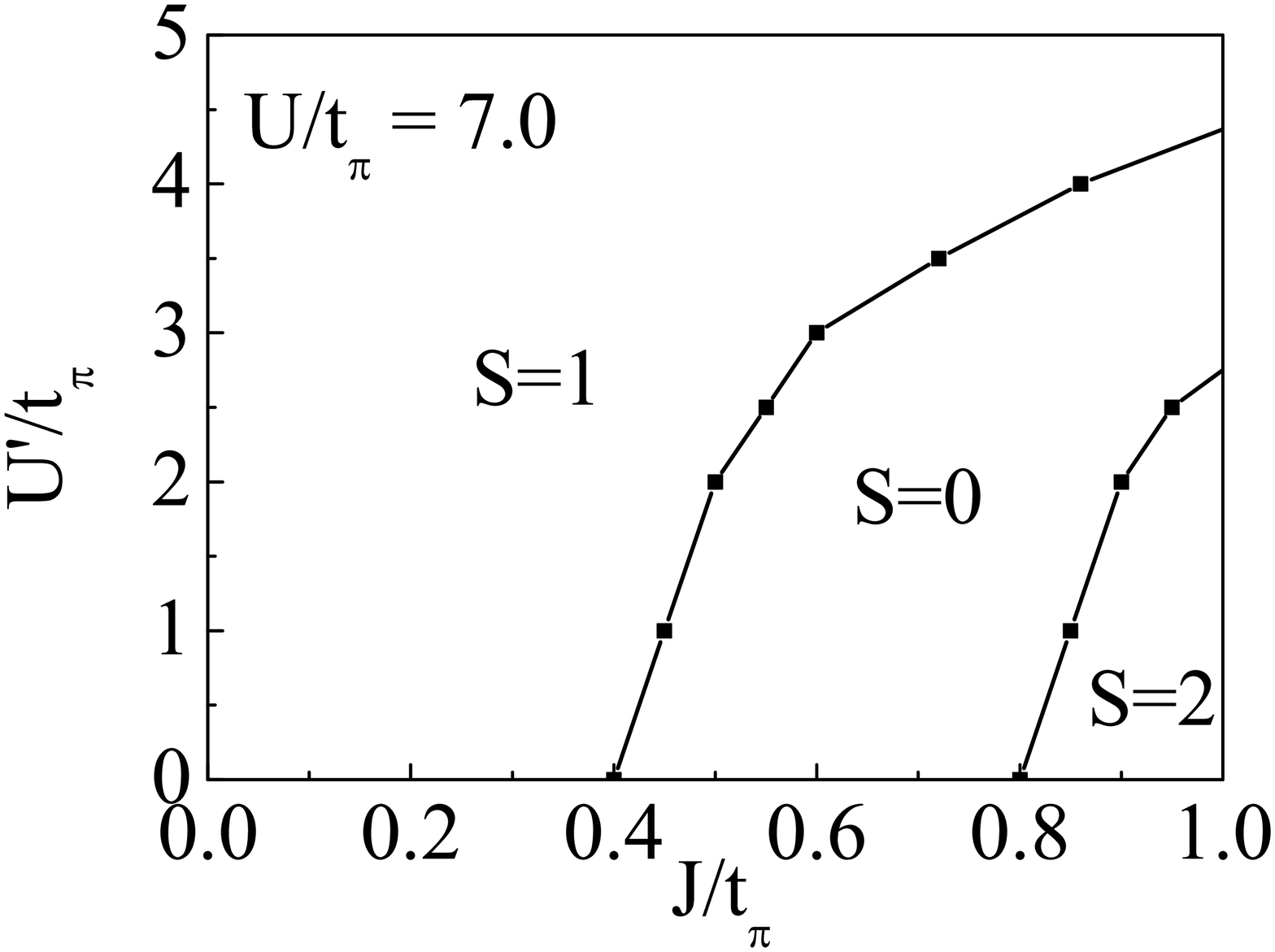}}
\caption{
Phase diagram in $U^\prime/t_{\pi}$ - $J/t_{\pi}$ plane for an isolated dianionic Bz, Bz$^{2-}$, system. 
}
\label{Fig1}
\end{center}
\end{figure}

In Fig. \ref{Fig1}, we show the phase diagram for the ground state in the $U^\prime/t_{\pi}$ - $J/t_{\pi}$ plane within manifolds of $S=2$, $S=1$, and $S=0$.
When $J/t_{\pi}$ is increased, the ground state changes from $S=1$ to $S=0$, and changes from $S=0$ to $S=2$.
The $S=2$ state is such that the spins of four holes become ferromagnetic by the ferromagnetic exchange coupling $J$. 
The characters of electronic states of the $S=1$ state and $S=0$ state are discussed later.

\begin{figure}[h]
\begin{center}
\rotatebox{0}{\includegraphics[width=1\linewidth]{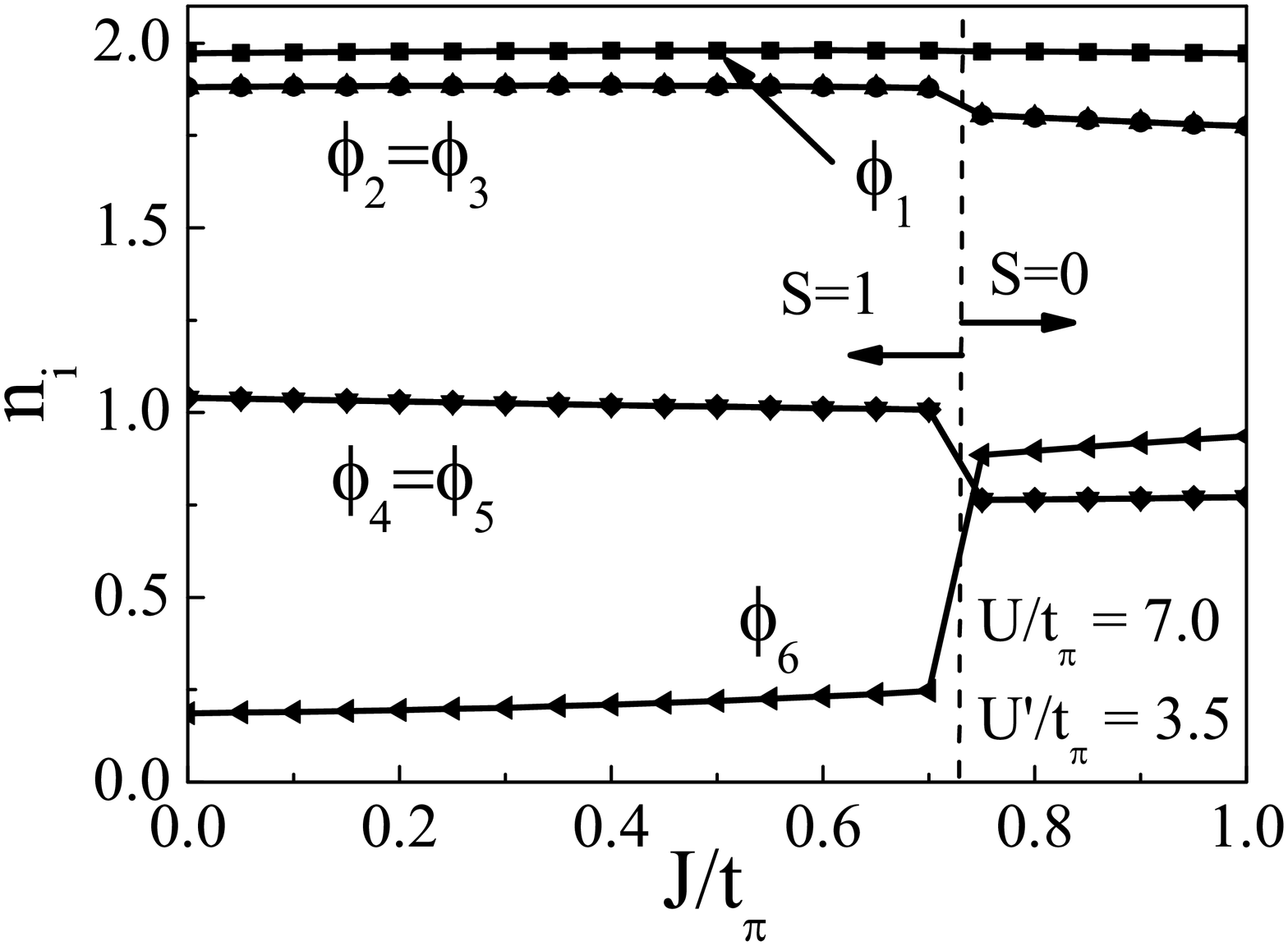}}
\caption{
Number of electrons of i-th molecular orbital $\phi_i$ (eq.(\ref{eq:1a}) $\sim$ eq.(\ref{eq:1f})) as a function of $J/t_{\pi}$. }
\label{Fig2}
\end{center}
\end{figure}

Figure \ref{Fig2} shows the number of electrons in the molecular orbital $\phi_{i}$ (eq.(\ref{eq:1a}) $\sim$ eq.(\ref{eq:1f})) as a function of $J/t_{\pi}$ at $U/t_{\pi}=7.0$ and $U'/t_{\pi}=3.5$.
For $J/t_{\pi} \lsim 0.7$, the number of electrons is evaluated to be $n_{\phi_{1}} \sim 2.0$, $n_{\phi_{2}}=n_{\phi_{3}} \sim 1.9$, $n_{\phi_{4}}=n_{\phi_{5}} \sim 1.0$, and $n_{\phi_{6}} \sim 0.2$, respectively.
On the other hand, for $J/t_{\pi} \gsim 0.7$, the number of electrons is evaluated to be $n_{\phi_{1}} \sim 2.0$, $n_{\phi_{2}}=n_{\phi_{3}} \sim 1.8$, $n_{\phi_{4}}=n_{\phi_{5}} \sim 0.8$, and $n_{\phi_{6}} \sim 0.9$, respectively.
When the ground state changes from the $S=1$ state to the $S=0$ state, the number of electrons in the $\phi_6$ orbital increases dramatically. 

\begin{figure}[h]
\begin{center}
\rotatebox{0}{\includegraphics[width=1\linewidth]{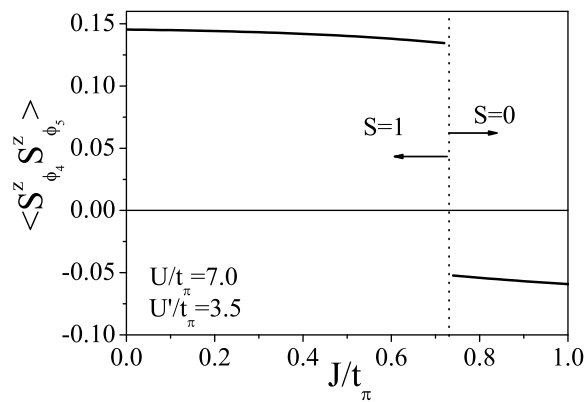}}
\caption{
Spin-spin correlation between LUMOs ($\phi_{4}$ and $\phi_{5}$) as a function of $J/t_{\pi}$.
The dotted line is the border of the $S=1$ state and the $S=0$ state. }
\label{Fig3}
\end{center}
\end{figure}

A spin-spin correlation between LUMOs is shown in Fig. \ref{Fig3}, where 
the spin-spin correlation is given by $\langle S_{\phi_{4}}^{z}S_{\phi_{5}}^{z} \rangle=\big[\langle n_{\phi_{4}\uparrow}n_{\phi_{5}\uparrow}\rangle -\langle n_{\phi_{4}\uparrow}n_{\phi_{5}\downarrow}\rangle -\langle n_{\phi_{4}\downarrow}n_{\phi_{5}\uparrow}\rangle+\langle n_{\phi_{4}\downarrow}n_{\phi_{5}\downarrow}\rangle \big]/4$.
For $J/t_{\pi} \lsim 0.7$, the correlation is $\langle S_{\phi_{4}}^{z}S_{\phi_{5}}^{z} \rangle \sim 0.15$, and it is ferromagnetic.
This correlation is understood as follows.
The number of electrons in each LUMOs is evaluated to be $n_{\phi_4}=n_{\phi_5} \sim 1$ as shown in Fig.\ \ref{Fig2}.
Two electrons occupying LUMOs form the ferromagnetic state due to the ferromagnetic interaction between two electrons on LUMOs (Hund's rule coupling). 
At $U^\prime/t_{\pi}=3.5$ and $J/t_{\pi}=0.3$, the ferromagnetic interaction between LUMOs, $J_{p}$, is estimated to be $J_p = E(S=0)-E(S=1) \sim 0.34t_{\pi}$.
Using $t_{\pi}$ $\sim$  2.5 eV~\cite{Benzene}, $J_p$ is of the order of $0.9$eV.
 On the other hand, we find that for $J/t_{\pi} \gsim 0.7$, the correlation is $\langle S_{\phi_{4}}^{z}S_{\phi_{5}}^{z} \rangle \sim -0.05$, and is antiferromagnetic, although the occupancy of $\phi_4$ and $\phi_5$ is $n_{\phi_4}=n_{\phi_5} \sim 1$, as in the case of the ferromagnetic state.
The reason why the spin-spin correlation becomes antiferromagnetic is due to the increase of the correlation between the $\phi_6$ orbital and other molecular orbitals.
For $U^\prime/t_{\pi}=3.5$, and $J/t_{\pi}=0.9$, the antiferromagnetic interaction between LUMOs, $J_{p}$, is estimated to be $J_p = E(S=0)-E(S=1) \sim -0.1t_{\pi} \sim -0.3$eV.

In conclusion, we have clarified the phase diagram of the ground state of Bz$^{2-}$ by the numerical diagonalization of the extended Hubbard model. 
We have shown that the ground state can be the spin-singlet state, depending on the size of ferromagnetic coupling between adjacent carbon atoms.
Then, we have found that the exchange interaction between two electrons on LUMOs can be antiferromagnetic, in contrast to the ferromagnetic exchange interaction expected by the Hund's rule coupling.


\begin{thebibliography}{99}
\bibitem{Murrell}  
J. N. Murrell and A. Hinchliffe:  
Molecular Physics. {\bf 11} (1966) 101. 
\bibitem{Jesse}  
R. E. Jesse and G. J. Hoytink:  
Chem. Phys. Lett. {\bf 1} (1967) 109.  
\bibitem{Nakayama}
M. Nakayama, H. Ishikawa, T. Nakano and O. Kikuchi:  
J. Mol. Struct. Theochem. {\bf 184} (1989) 369.
\bibitem{Hirsch}
J. E. Hirsch: Phys. Rev. B {\bf 40} (1989) 2354, {\bf 40} (1989) 9061.
\bibitem{Tang}
S. Tang and J. E. Hirsch:  
Phys. Rev. B {\bf 42} (1990) 771. 
\bibitem{McWeeny}
R. McWeeny: Proc. Roy. Soc. (London) A {\bf 227} (1955) 288. 
\bibitem{Benzene}
T. Yonezawa, T. Nagata, H. Kato, A. Imamura and K. Morokuma: 
{\it Introduction to Quantum Chemistry, Part I}, 3rd ed. 
(Kagaku-Dojin, Kyoto, 1983).
\end{thebibliography}
\end{document}